\documentclass[reprint,amsmath,amssymb,aps]{revtex4-1}
\usepackage{graphicx}
\usepackage{epstopdf}

\usepackage{color}
\begin{document}
 \title{Critical-point behavior of a measurement-based quantum heat engine}
\author{Suman Chand}
 \email{suman.chand@iitrpr.ac.in}
\author{Asoka Biswas}%
\affiliation{%
Department of Physics, Indian Institute of Technology Ropar, Rupnagar, Punjab 140001, India
}%
\date{\today}%

\begin{abstract}
We study how  a quantum heat engine performs across the critical value of an external parameter, pertaining to the quantum phase transition. Considering a two-ion system subjected to a magnetic field, we show that the system performs in a quantum Otto cycle above a critical value of the magnetic field, while below such critical point, it does not operate in a heat cycle at all.  Moreover, at the critical point, its interaction with an ancillary ion deteriorates the performance of the system as a heat engine. We further show that a strong interaction between the constituent ions of an ion-based system is crucial for it to work in a heat-work cycle, while  the coupling to the ancillary system must be minimized.  
\end{abstract}

\maketitle

\section{\label{sec:i}Introduction}
A heat engine operates between two heat reservoirs, which are at thermal equilibrium at two different temperatures $(T_C, T_H > T_C)$, and employs some of the absorbed heat into delivering certain work. The efficiency of such an engine is limited by the so-called Carnot's limit \cite{Kerson Huang book on Statistical Mechanics} $\eta=1-T_C/T_H$. The working fluid in a standard heat engine can be a gas of particles or some liquid. However, with respect to recent thrust on research on nano-dimensional systems, it becomes quite contextual, to study the behavior of heat engine, if the working fluid consists of a few particles. This naturally invokes quantum mechanical aspects in the study of heat engines \cite{Scovil maser paper}, leading to a new perspective into thermodynamics \cite{Thermodynamics book G. Mahler,The Role of Quantum Information in Thermodynamics—A Topical Review-J. Goold,Quantum Thermodynamics-S. Vinjanampathy,Introduction to Quantum Thermodynamics: History and Prospect- R. Kosloff}  and many possibilities of heat management at the quantum level. For example, quantum systems like spins and other two-level systems  \cite{Quan-Thermodynamics cycle,kosloff,Quantum Otto engine of a two-level atom with single-mode fields,Entangled QHE based on two two spin systems with DM,Four-level entangled quantum heat engines,Thermal entanglement in two-atom cavity QED and the entangled quantum Otto engine,johal,Effects of reservoir squeezing on quantum systems and work extraction,Atlintas,Thermal entangled four level QOE,A Special entangled QHE based on two qubit Heisenberg XX model,Special coupled quantum otto cycle,Thermal entangled QHE working with 3-qubit XX model,Lipkin-Meshkov-Glick Model in QOE,Performence of coupled system as QHE}, quantum harmonic oscillators \cite{Quan-Thermodynamics cycle,Performence of coupled system as QHE,Quantum harmonic oscillator Kosloff,Efficiency at maximum power of a quantum heat engine based on two coupled oscillators}, cavity QED \cite{Quantum-classical transition of photon-Carnot engine induced by quantum decoherence-Quan}, single trapped ion \cite{single ion-Abah}, optomechanical systems \cite{optomechanical system-Meystre}, quantum dots \cite{Magnon-driven quantum-dot heat engine}, cold bosons \cite{Isolated quantum heat engine- bosonic system} etc. can be employed to operate as heat engines. 

More importantly, a system with initial coherence can operate with an efficiency beyond Carnot's limit \cite{Scully Science}, in presence of a heat bath with nonclassical properties. In the presence of inter-particle interaction, entanglement and nonclassical correlations between the particles in such systems arise that can substantially enhance the engine efficiency \cite{kosloff,Entangled QHE based on two two spin systems with DM,Four-level entangled quantum heat engines,Thermal entanglement in two-atom cavity QED and the entangled quantum Otto engine,johal,Effects of reservoir squeezing on quantum systems and work extraction,Atlintas,Thermal entangled four level QOE,A Special entangled QHE based on two qubit Heisenberg XX model,Special coupled quantum otto cycle,Thermal entangled QHE working with 3-qubit XX model,Lipkin-Meshkov-Glick Model in QOE,Performence of coupled system as QHE,Quantum Brayton cycle with coupled systems as working substance}. Note that such a long-range correlation can appear during quantum phase transition \cite{S. Sachdev book- Quantum Phase Transition} at critical points. In this regard, it becomes imperative to explore the behavior of quantum heat engine at the quantum critical point. 

Quantum phase transitions (QPT) correspond to transition from one ground state to the other at a critical value of an external control parameter at absolute zero temperature \cite{S. Sachdev book- Quantum Phase Transition}. It refers to a level crossing and nonanalyticity of the ground-state energy at this quantum critical point. The correlation between the particles at criticality exhibits long-range behavior, referring to a strongly coupled many-body system. It was recently shown in a Lipkin-Meshkov-Glick model \cite{Quantum thermodynamic cycle with quantum phase transition}  that efficiency of the quantum heat engine is enhanced at the critical point. This suggests that long-range correlations (namely, entanglement) may be responsible for enhancement of the efficiency. To further explore whether this is a generic feature at criticality, we consider, in this paper, a system of two trapped ions as a working fluid and show that its behavior as a heat engine is different across the critical point. Moreover, while the interaction between these two ions enhances the engine efficiency, their coupling to a third ion (as a part of the spin chain) has an adverse effect on the efficiency at the critical point. In fact, we conjecture that the internal ion-ion correlation and the external control parameter exhibit a cumulative effect on the efficiency, which may not be interpreted in terms of entanglement among the ions. Rather, the nature of the interaction governs the performance of the engine.  Note that we consider each trapped ion confined to its two lowest electronic energy states in the Lamb-Dicke limit \cite{blatt} and to two lowest vibrational eigenstates (with average phonon number much less than unity). This corresponds to a temperature of the order a few nano-Kelvin and may be considered a practical approximation to absolute zero temperature, as required in QPT by definition.

The structure of the paper is as follows. In Sec. II, we describe the model and our main results, including the engine operation at the critical points and the effect of a third qubit. In Sec. III, we conclude the paper with an outlook. In the Appendix, we discuss the uniques features of our model, along with relevant analysis. 

\section{Implementation of the heat cycles}
\subsection{Our model}\label{s:ii}
We start with three trapped ions in a one-dimensional array, each with its lowest-lying electronic states $|\pm\rangle$ as the relevant energy levels.  These ions share a common vibrational mode $a$. 
The Heisenberg XX-type interaction among these ions and the interaction between the vibrational mode and the electronic modes can be described by the following Hamiltonian  (in unit of Planck's constant $\ensuremath{\hbar=1}$):
\begin{equation}
H = H^{(0)}_{12}+H^{(0)}_{3}+H^{\rm int}_{12}+H^{\rm int}_{23}+H_{{\rm ph}}+H_{12,{\rm ph}}+H_{3,{\rm ph}}\;,
\end{equation}
where
\begin{eqnarray}
H^{(0)}_{12} & = & B_{1}\sigma_{z}^{(1)}+B_{2}\sigma_{z}^{(2)}\;;\;\; H^{(0)}_{3}=B_{3}\sigma_{z}^{(3)}\nonumber\\
H^{\rm int}_{12}&=&J_{1}\left(\sigma_{+}^{(1)}\sigma_{-}^{(2)}+\sigma_{-}^{(1)}\sigma_{+}^{(2)}\right) \nonumber \;;\\
H^{\rm int}_{23}& = & J_{2}\left(\sigma_{+}^{(2)}\sigma_{-}^{(3)}+\sigma_{-}^{(2)}\sigma_{+}^{(3)}\right)\;, \nonumber \\
 H_{\rm ph} & = & \omega a^{\dagger}a\;, \nonumber \\
H_{12,{\rm ph}} & = & k_{1}\left(a^{\dagger}\sigma_{-}^{(1)}+\sigma_{+}^{(1)}a\right)+k_{2}\left(a^{\dagger}\sigma_{-}^{(2)}+\sigma_{+}^{(2)}a\right)\nonumber \\ 
H_{3,{\rm ph}}&= & k_{3}\left(a^{\dagger}\sigma_{-}^{(3)}+\sigma_{+}^{(3)}a\right)\;.
\end{eqnarray}
Here $H^{(0)}_{12}$ represents the unperturbed Hamiltonian of two ions 1 and 2, which interact with each other with the corresponding coupling constants $J_{1}$, $J_{2}$ represents the strength of  interaction between the second and the third ion), $H_{\rm ph}$ is the energy of the vibrational mode with frequency $\omega$, and  $H_{12,{\rm ph}}$ and $H_{3,{\rm ph}}$ define the interaction between the internal and the vibrational degrees of freedom of the respective ion. The interaction strength between the electronic transitions of the $l$th ion and the vibrational mode is given by $k_{l}=k$ ($l\in 1,2,3$). The 
magnetic field of strength $B_{l}$ is applied along the quantization axis to the $l$th ion. The cases $J_i>0$ and $J_i<0$ ($i\in 1,2$) correspond to the antiferromagnetic and the ferromagnetic interactions, respectively. In this paper, we choose the antiferromagnetic case only.

In our model, we use the third ion as an auxiliary system and consider the joint electronic degrees of freedom of the ions 1 and 2 as the working substance S of the heat engine.
Therefore the unperturbed system Hamiltonian can be identified as 
\begin{eqnarray}
H_S = H^{(0)}_{12} + H^{\rm int}_{12}\;.
\label{systemH}
\end{eqnarray}
The eigenvalues of this Hamiltonian $H_S$ are given by (considering $B_1=B_2=B$)
\begin{equation}
E_{1}=-2B,\, E_{2}=2B,\, E_{3}=-J_{1},\, E_{4}=+J_{1}\,,
\label{eigenvalues}
\end{equation}
 with the respective eigenstates $|E_1\rangle=|--\rangle$, $|E_2\rangle=|++\rangle$, $|E_3\rangle= \frac{1}{\sqrt{2}}\left(\left|-+\right\rangle -\left|+-\right\rangle \right)$ and $|E_4\rangle=\frac{1}{\sqrt{2}}\left(\left|-+\right\rangle +\left|+-\right\rangle \right)$.

As described by the Hamiltonian $H$, the system $S$ is subjected to the interaction with a third ion and the vibrational mode $a$. While the mode $a$ can be chosen as a part of the heat cycles, the third ion rather influences the correlation into the system $S$ and therefore the efficiency of the heat engine. In the following we will first briefly describe the operation of the heat engine and then the effect of the third ion.
 
As described in the Introduction, the vibrational mode can be considered a two-level {\it cold} bath (with the relevant phonon-number states $|0\rangle$ and $|1\rangle$) with an average phonon number $\bar{n}_{\rm ph}\ll 1$. For instance, a single Be ion can be cooled utilizing standard ion trapping procedure, to such an extent that the average motional quantum number can be of the order of 0.02 (see, for example, Ref. \cite{Monroe ion trap}). We here emphasize that a finite-level system will act as a bath, as coupling to such a bath frequently prompts decoherence of the system (see, e.g., Ref. \cite{Mam and Brummer paper}. The system $S$ and the ion 3 continuously interact with this effective cold bath through the Hamiltonian $H_{12,{\rm ph}}$ and $H_{3,{\rm ph}}$, respectively, while the thermal environment at an equilibrium temperature $T_H$ simultaneously interacts with the system $S$, ion 3, and the vibrational mode.

In this paper, we focus on the quantum Otto cycle, which consists of two quantum isochoric and two quantum adiabatic stages. Such an engine resembles more realistic situations than idealistic Carnot engines, with a reciprocating heat cycle such that allows energetic changes to be distinguished with separate stages. This means that in each stage, either work is extracted from (or done on) the system, or thermal energy is exchanged between the system and the reservoirs \cite{strong-coupling} but not both. First, in the isochoric heating stage ($1\rightarrow 2$, Fig. 1), the ions get thermalized to an equilibrium temperature $T_H$ of the ambient hot bath. To estimate the heat exchanged by the system with the bath during this stage, we start with the joint basis $|\alpha\beta\gamma j\rangle$, where $\alpha,\beta,\gamma\in \pm$ represent the electronic states of the ions 1, 2, and 3, respectively, and $j\in 0,1$ the states of the mode $a$. In this basis, the eigenstates of the Hamiltonian $H$ can be represented as $|U_n\rangle=\sum_{\alpha,\beta,\gamma,j}a_{\alpha\beta\gamma j}^n|\alpha\beta\gamma j\rangle$, where $n\in [1,16]$ and $a_{\alpha\beta\gamma j}^n$ is the probability amplitude of the corresponding basis states in the $n$th eigenstate. The interaction with the thermal bath leads to the following mixed state of the joint system:
\begin{equation}
\rho_1^{(H)}=\sum_{n=1}^{16}p_n|U_n\rangle\langle U_n|\;,\;p_n=\frac{\exp\left(-U_n/k_BT_H\right)}{\sum_{n=1}^{16}\exp\left(-U_n/k_BT_H\right)}\;,
\label{rho1H}
\end{equation}
where $p_n$ is the occupation probability of the $n$th eigenstate $|U_n\rangle$ (with corresponding eigenvalue $U_n$) and $k_{B}$ is the Boltzmann constant. Note that this state is achieved at the steady state irrespective of the initial preparation of the ions.

As we are interested in calculating the heat exchanged by the two-ion system $S$ during this stage, we next obtain the reduced density matrix of the same in the basis $\{|E_i\rangle\}$, by taking the partial trace over the vibrational states and the ion 3. The average energy of the system can be written as $U=\sum_{i=1}^4 P_i E_i$, where $P_i$ is the occupation probability of the state $|E_i\rangle$. If the initial (final) probability for being in the $i$th eigenstate $|E_i\rangle$ is $P_{i}\left(T_{L}\right)$ [$P_{i}\left(T_{H}\right)$], then the heat exchanged with the hot bath by the system $S$ during this stage is given by
\begin{equation}
Q_{H}=\sum_{i=1}^{4}E_{i}^H\left\{ P_{i}\left(T_{H}\right)-P_{i}\left(T_{L}\right)\right\}\;. 
\end{equation}
Note that this clearly depends upon the initial preparation of the ion,
During this process, the magnetic field is kept fixed at $B_{l}=B_{H}$ ($l\in, 1,2,3$), such that the corresponding eigenvalues $E_{i}^{H}$ of the system Hamiltonian $H_S$ also remain constant and therefore no work is done. Due to the change in the occupation probabilities, only the heat is exchanged during this cycle.

\begin{figure}
\includegraphics[width=8.5cm,height=5cm]{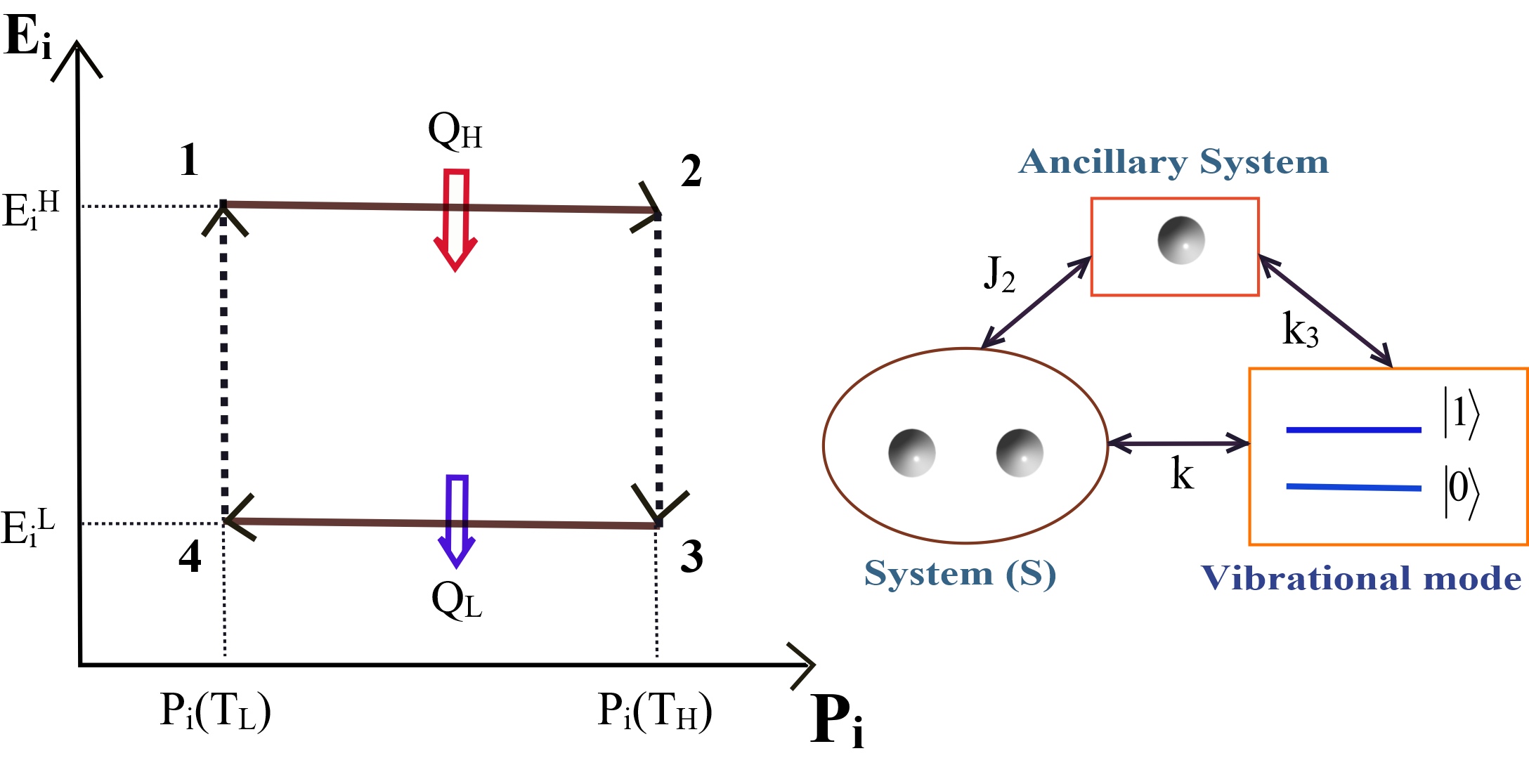}
\caption{(Color online) A schematic $E_{i}$-$P_{i}$ (energy levels vs occupation probabilities) diagram of a quantum Otto cycle. The solid (dashed) lines refer to the isochoric (adiabatic) processes. The inset displays the different interaction channels among the system $S$ (comprising two ions), an ancillary system (the ion 3), and the phonon mode, and the corresponding interaction strength.}
\end{figure}

In the adiabatic expansion stage ($2\rightarrow 3$, Fig. 1), the magnetic field $B_{l}$ ($l\in, 1,2$) is varied slow enough from $B_H$ to a smaller value $B_L$, such that the process remains adiabatic. This corresponds to a change in the eigenvalues  $E_{1,2}$ from $\mp 2B_H$ to $\mp 2B_L$, yet keeping the occupation probability of each eigenstate $\left|E_{i}\right\rangle$ nearly the same. Thus, there is no heat exchange between the system S and its environment, except certain work 
\begin{equation}
W_{1}=\sum_{i=1}^{4}P_{i}\left(T_{H}\right)\left(E_{i}^{L}-E_{i}^{H}\right)\;.
\end{equation}
For adiabatic evolution, a linear ramp such as $B(t)=B_H-(B_H-B_L)t/\tau$ could be chosen such that the adiabatic evolution takes place at a time $\tau$, much smaller than the thermalization time-scale $1/\gamma$ in which the heat bath (characterized by the temperature $T_{H}$) would become effective ($\gamma$ is the decay rate of the system $S$) (a similar condition is also considered in Ref. \cite{Single-Atom Heat Machines Enabled by Energy Quantization}). Therefore, the system remains nearly unaffected by the heat bath and effectively evolves as a closed system. So, during this stage, no heat is exchanged, and the change in entropy is zero. This refers to a reversible adiabatic process (isentropic process), that is associated with no internal friction or heat leak \cite{Quantum four-stroke heat engine Thermodynamics observable in a model with intrinsic friction, Irreversible Work and Inner Friction in Quantum Thermodynamics Process, Friction due to inhomogeneous driving of coupled spins in a quantum heat engine, Irreversible work and internal friction in a quantum Otto cycle of a single arbitrary spin}.  Moreover, such heat leaks would have appeared in presence of inhomogeneous magnetic field \cite{ Friction due to inhomogeneous driving of coupled spins in a quantum heat engine}, which is not the case in the present model.


Next, during the isochoric cooling stage ($3\rightarrow 4$, Fig. 1) of an Otto cycle, some amount of heat $Q_{L}$ is transferred from the system to the cold bath, while the magnetic field is maintained at $B_L$. In the present case, the vibrational mode $a$ is modelled as the cold bath and the heat release is performed by measuring the system $S$ and the ion 3 in a suitable basis (see  Refs. \cite{Our Paper- Single ion, Our Paper- Two ions} for details). The initial state $\rho_{2}^{\left(L\right)}$ for this stage (as obtained after the adiabatic process) can be written in the joint basis of the electronic states of 3 ions and the vibrational mode as
\begin{eqnarray}
\rho_{2}^{(L)} = \sum_{q,r =1}^{16}\rho_{2}^{\left(qr\right)}\left|q\right\rangle \left\langle r\right|\;, \end{eqnarray}
where $|q\rangle , |r\rangle\equiv |\alpha\beta\gamma j\rangle$, as defined before. Heat release from the system $S$ is equivalent to cooling it down to the ground state. So, a measurement in the ground state [e.g., in the state $|--\rangle$, when $J_1<2B_L$, see (\ref{eigenvalues})] would effectively mimic the isochoric cooling process, and the associated heat release can be expressed as 
\begin{equation}\label{ql1}
Q_L=\sum_{i=1}^4E_i^L[P_i(T_L)-P_i(T_H)]\;, 
\end{equation}
where $P_i(T_L)=1$ when $|E_i\rangle$ is the basis state in which the system $S$ is measured (in this case, the ground state) and $P_i(T_L)=0$ otherwise. 

In the last stage in the cycle (the adiabatic compression process, $4\rightarrow 1$, Fig. 1), the magnetic field is adiabatically restored to the value $B_H$ from $B_L$, such that the occupation probabilities of the energy eigenstates $|E_i\rangle$ of the system remain maintained at the values $P_{i}(T_{L})$.  The work done by the system during this stage is given by 
\begin{equation}\label{w2}
W_2=\sum_{i=1}^4P_i(T_L)(E_i^{H}-E_i^{L})\;.
\end{equation}

\subsection{Efficiency of the heat engine around the critical point}

%


\begin{figure}[!h]
$\begin{array}{cc}
 \includegraphics[width=0.49\linewidth]
 {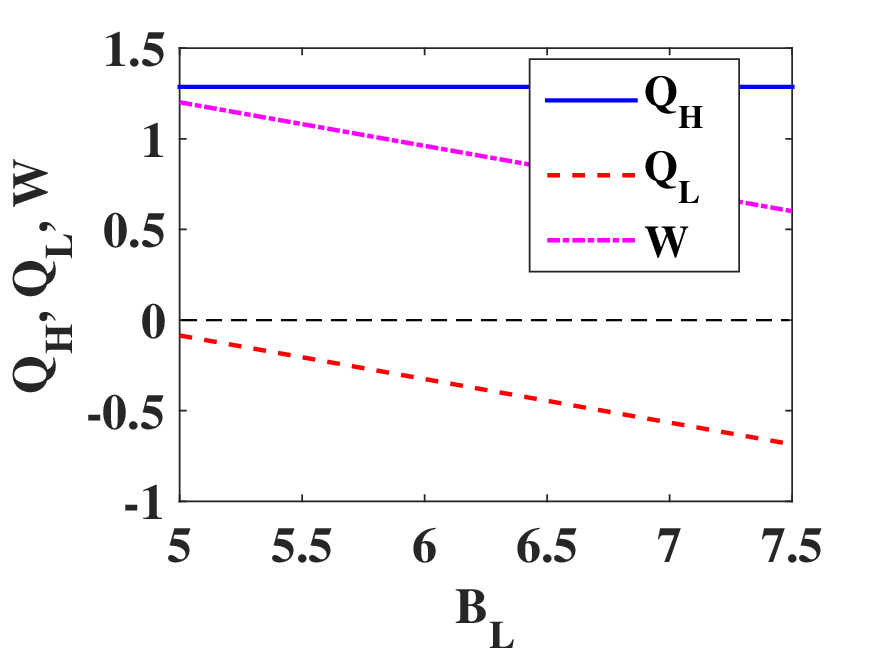}&
\includegraphics[width=0.49\linewidth]{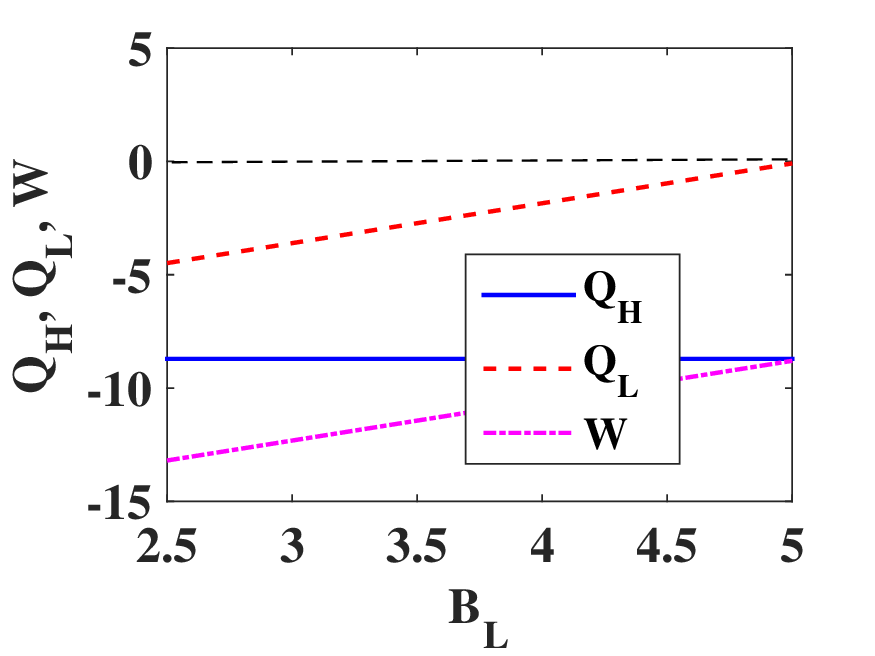}\\(a) & (b)
\end{array}$ 
\caption{(Color online) Variation of heat-exchanged $Q_H$ (solid blue) and $Q_L$ (dotted red), with the hot and the cold bath, respectively and the net work done $W$ (dot-dashed magenta)  as a function of the magnetic field  $B_{L}$ during exhaust stage, when the measurement is done in (a) the $|E_{1}\rangle$ state and (b) the $|E_{3}\rangle$ state.  The others parameters for the cycle are $B_H = 10, k=0.1, \omega =1, k_BT_H=3.5,$ and $J_1=J_{2}=10$. The physically acceptable parameter region for the engine to operate is obtained when the measurement is performed in the $|E_{1}\rangle$ state.}
\label{Fig: QH_QL_W___vs___BL____P1}
\end{figure}

As mentioned in the Introduction, the quantum phase transition occurs in a quantum system when the external driving parameter sweeps through a critical value. It corresponds to a nonanalyticity of the ground-state energy as one varies the parameter across the "critical point". Let us consider the dynamics of a quantum system under the action of the Hamiltonian $H\left(\lambda\right)=H_{0}+\lambda H_{1}$. If we change the parameter $\lambda$, then there can be a level crossing (crossing of two eigenvalues at a critical value $\lambda=\lambda_0$) at which an excited state turns into a lowest-energy state (i.e., the ground state). Such a form of non-analyticity (i.e., the level crossing) corresponds to the first-order QPTs \cite{S. Sachdev book- Quantum Phase Transition,Work and quantum phase transitions: Quantum latency}. Despite the fact that absolute zero is not achievable, it is logical to consider the presently achievable low-temperature domain \cite{Quantum resources for purification and cooling: fundamental limits and opportunities, Cooling in strongly correlated optical lattices: prospects and challenges} to closely match with the parameter domain that is required in QPTs, by definition. It is also assumed that the ground state is nondegenerate \cite{Work and quantum phase transitions: Quantum latency}.

\begin{figure}[!h]
$\begin{array}{cc}
 \includegraphics[width=0.49\linewidth]{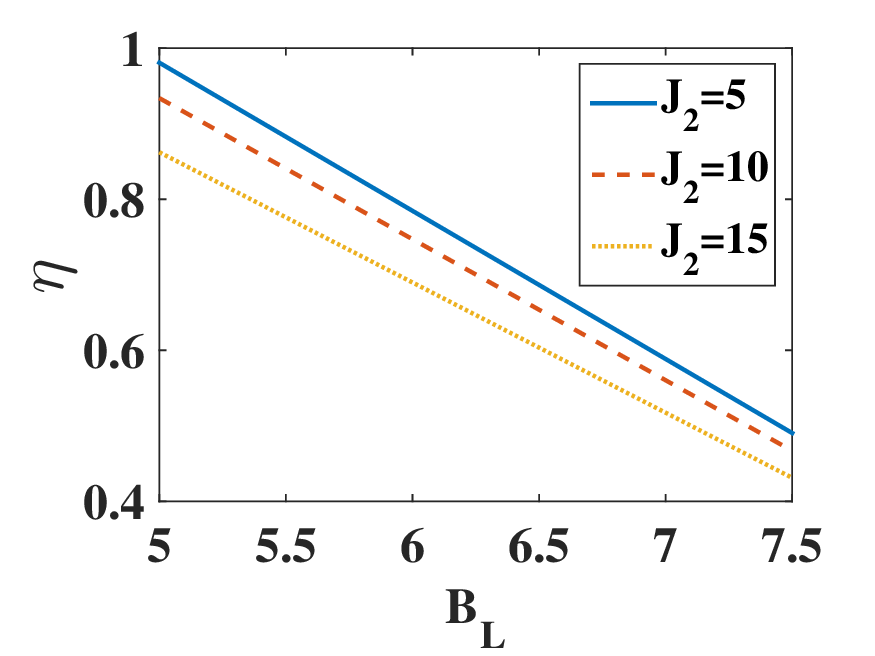}&
\includegraphics[width=0.49\linewidth]{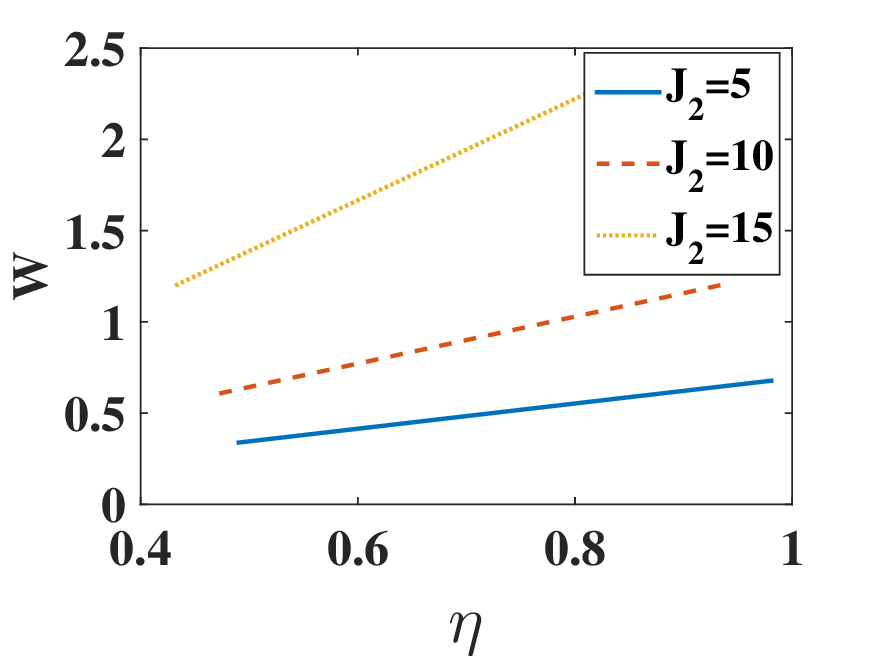}\\(a) & (b)
\end{array}$ 
\caption{(Color online)  (a) Variation of the efficiency $\eta$ as a function of the magnetic field  $B_{L}$ and (b) the work $W$ done by the system with the efficiency $\eta$, for different values of $J_2$, and for the same range of $B_L$ as in (a). The system is measured in $|E_1\rangle$ state. The other parameters are the same as in Fig. \ref{Fig: QH_QL_W___vs___BL____P1}.}
\label{Fig: eta__and__W___vs___BL____P1}
\end{figure}

In the present case, as one sweeps the magnetic field $B$ (equivalent to the control parameter $\lambda$, as discussed above), the eigenvalues of the Hamiltonian (\ref{systemH}) display a level crossing at $B=J_1/2$, referring to a critical point. Below, we first study the behavior of the heat engine above and below the critical point. 

\begin{figure}[!h]
$\begin{array}{cc}
 \includegraphics[width=0.49\linewidth]{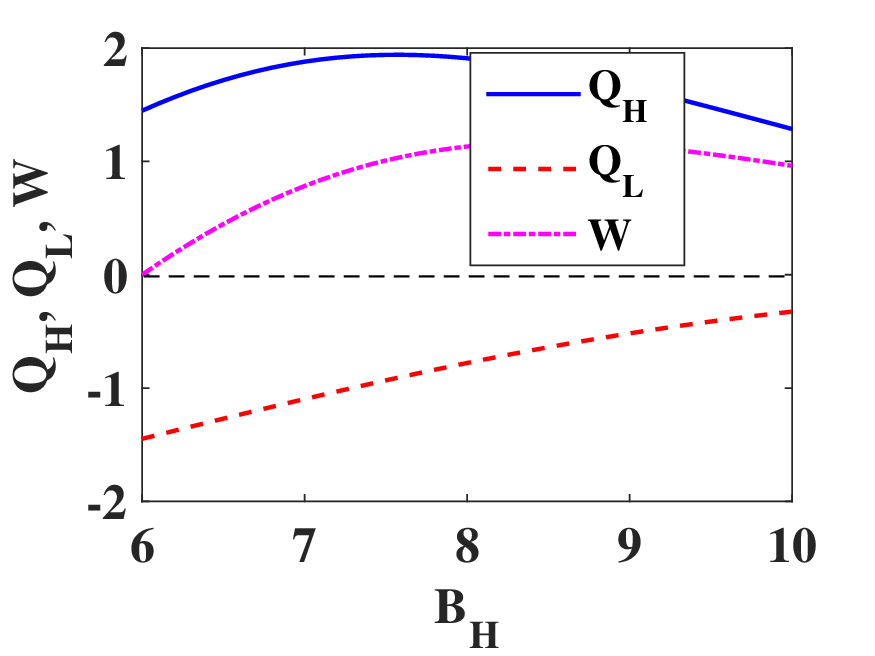}&
\includegraphics[width=0.49\linewidth]{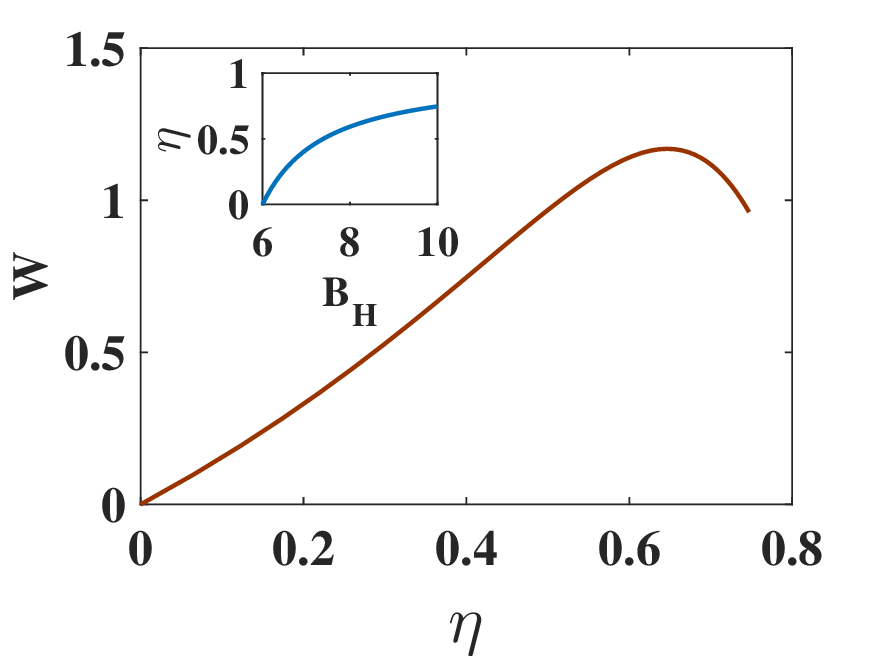}\\(a) & (b)
\end{array}$ 
\caption{(Color online) (a) Variation of heat-exchanged $Q_H$ (solid blue) and $Q_L$ (dotted red), with the hot and the cold bath, respectively and the net work done $W$ (dot-dashed magenta)  as a function of the magnetic field  $B_{H}$ during isochoric heating step and (b) the work $W$ done by the system with the efficiency $\eta$, while the system is measured in  $|E_1\rangle$ state. We have chosen $B_{L}=6$, a value above the critical point. The other parameters are the same as in Fig. \ref{Fig: QH_QL_W___vs___BL____P1}. The inset in (b) shows the variation of $\eta$ with $B_{H}$.}
\label{Fig: Q_H Q_L W vs BH for BL__6___ J1__10_____J2__10}
\end{figure}

As one decreases the magnetic field from $B_H$ to $B_L$, such that $B_L$ remains larger than $J_1/2$ (above the critical point), the state $|E_1\rangle=|--\rangle$ remains the ground state of the system. So, a measurement in this state during the isochoric cooling process cools down the system and the reminiscent heat gets absorbed by the ancillary systems. In this case, the heat absorbed $Q_H$ becomes positive, while $Q_L<0$ and the work done becomes positive [see Fig.\ref{Fig: QH_QL_W___vs___BL____P1}(a)]. This refers to execution of a heat engine. Note that as $B_L$ tends to the critical value $J_1/2$, the efficiency of this heat engine, $\eta=\frac{\rm Work\,Output}{\rm Heat\,Input}=\frac{Q_{H}+Q_{L}}{Q_{H}}$, tends to its maximum value [see Fig. \ref{Fig: eta__and__W___vs___BL____P1}(a)]. We further show in Fig. \ref{Fig: eta__and__W___vs___BL____P1}(b) how the work $W$ done by the system varies with the efficiency $\eta$. To obtain this variation we have changed the magnetic field $B_{L}$ above the critical point, i.e., $B_L\ge J_1/2$, while the largest value of $\eta$ refers to the the critical point $B_L=J_1/2$. It is clear from this figure that, close to the critical point, both the work output and the efficiency of the heat engine are large. Such a monotonic behavior of work-efficiency plot has been reported also in Ref. \cite{The power of a critical heat engine}. With the change in $B_{L}$, the work done by the system changes, while the heat absorbed $Q_H$ remains the same, leading to a linear work-efficiency behavior (see Appendix for a relevant analysis). Instead, if one changes $B_{H}$, keeping $B_{L}$ fixed at a value above the critical point, the $Q_{H}$ gets modified, leading to a parabolic nature in the work-efficiency dependence (see Fig. \ref{Fig: Q_H Q_L W vs BH for BL__6___ J1__10_____J2__10}). This kind of variation has been shown also in Refs. \cite{Generalized model and optimum performance of an irreversible quantum brayton engine with spin system,gordon}.
\begin{figure}[!h]
 \includegraphics[width=0.49\linewidth]{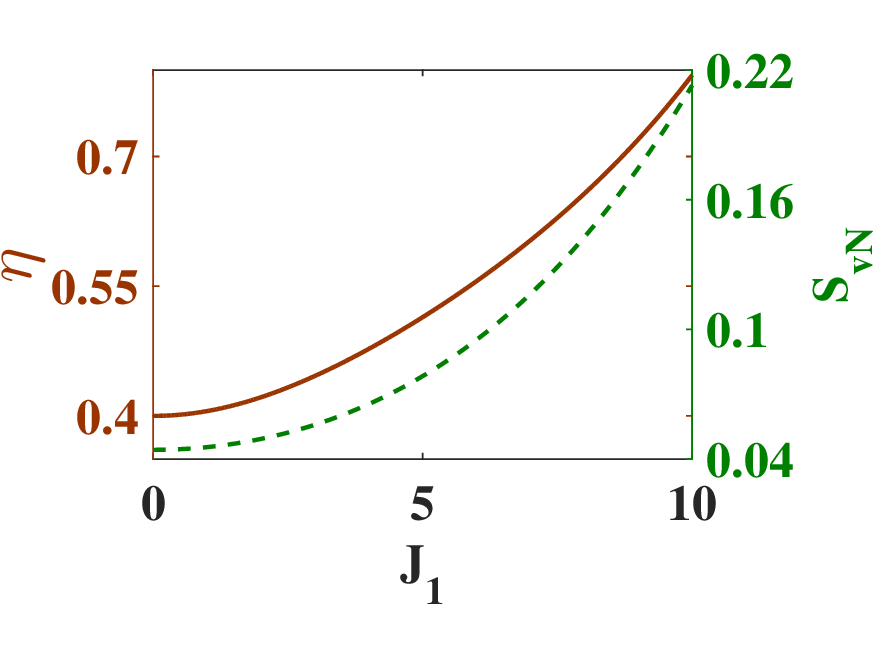}
\caption{(Color online)  Variation of the efficiency $\eta$ (solid red) and von Neumann entropy $S_{vN}$ (dashed green) as a function of $J_{1}$ for $J_{2}=0.1$. We have chosen $B_L=6$, a value above the critical point for the variation of $J_{1}$ and the other parameters are the same as in Fig. \ref{Fig: QH_QL_W___vs___BL____P1}. The system is measured in the $|E_1\rangle$  state.}
\label{Fig: eta__S__vs___J1__P1}
\end{figure}
On the other hand, below the critical point, the system behaves quite differently. When the magnetic field is decreased to a value below the critical value $J_1/2$, the state $|E_1\rangle=|--\rangle$ no longer remains the ground state but rather the state $|E_3\rangle=\sqrt{1/2}(|-+\rangle-|+-\rangle)$ becomes the minimum energy state. So a measurement in this state could lead to cooling in the isochoric cooling stage, $3\rightarrow 4$, i.e., $Q_L<0$. On the contrary, we find that $Q_H$ becomes negative, referring to heat release to the hot bath during the isochoric heating stage, $1\rightarrow 2$. This could otherwise lead to a refrigeration effect, if $Q_L>0$ and $W<0$ (i.e., if heat would be absorbed from the cold bath, at the expense of work done {\it on} the system). In our case, however, all the heat and work terms, i.e., $Q_H$, $Q_L$, and  $W$ remain negative [see Fig.\ref{Fig: QH_QL_W___vs___BL____P1}(b)]. This represents an unphysical situation, which corresponds neither to a heat engine cycle nor a refrigeration effect. 

This clearly indicates that the critical point $B_L=J_1/2$ defines the lower limit of the magnetic field till which one can extract certain work from the system. By reversing the direction of the magnetic field, one can attain a similar boundary point, namely, $B_L=-J_1/2$, at which the eigenvalues of (\ref{systemH}) display a level crossing between the states $|E_2\rangle$ and $|E_3\rangle$. 

We next study how the interaction between two spins affects the efficiency $\eta$ above the critical point. We relate this to the von Neumann entropy of the system $S$. The von Neumann entropy \cite{Von Neumann book} $S_{vN}\left(\rho\right)=-k_{B}{\rm Tr}\left(\rho\ln\rho\right)$ is known to provide a signature of quantum correlations, and particularly, for a two-qubit system (as in our case of two-ion system $S$) quantifies the entanglement. For a two-qubit system, $\rho$ is the reduced density matrix of one of the qubits, obtained by partial trace over the Hilbert space of the other qubit. We find that both the von Neumann entropy $S_{vN}$ at the end of the isochoric heating  stage and the efficiency $\eta$ increase with  $J_{1}$ (see Fig. \ref{Fig: eta__S__vs___J1__P1}). For a fixed value of $J_2$, an increase in  $J_{1}$ leads to increase in the correlation (or, more precisely speaking, the entropic entanglement) between the ion 1 and the ion 2, and so the efficiency. Note that we have calculated the entropy of the state at the end of the isochoric heating process, when the system attains a thermal equilibrium. This can also be treated as the entropy change during heating, as the initial state before thermalisation is a pure state, thanks to the projective measurement in a pure state during the stage $3\rightarrow 4$, followed by the adiabatic isentropic stage $4\rightarrow 1$. We emphasize that the total entropy change during the full cycle is zero in the present model, as the entropy
is a state function and the system returns to the initial state after
one complete cycle, and therefore, no heat leak occurs into or out of the system.


\subsection{Engine efficiency at the critical point}
\begin{figure}[!h]
$\begin{array}{cc}
 \includegraphics[width=0.49\linewidth]{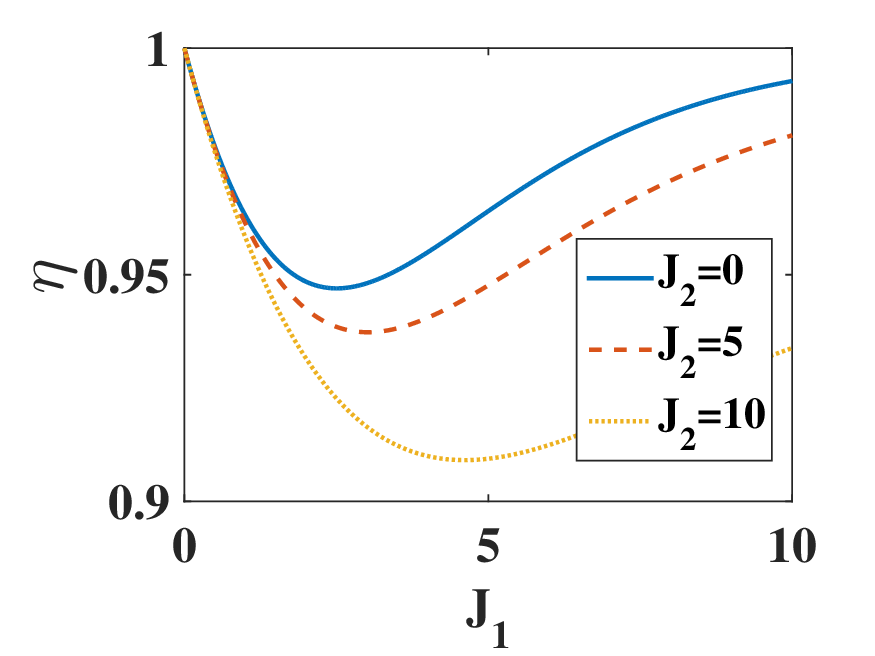}&
\includegraphics[width=0.49\linewidth]{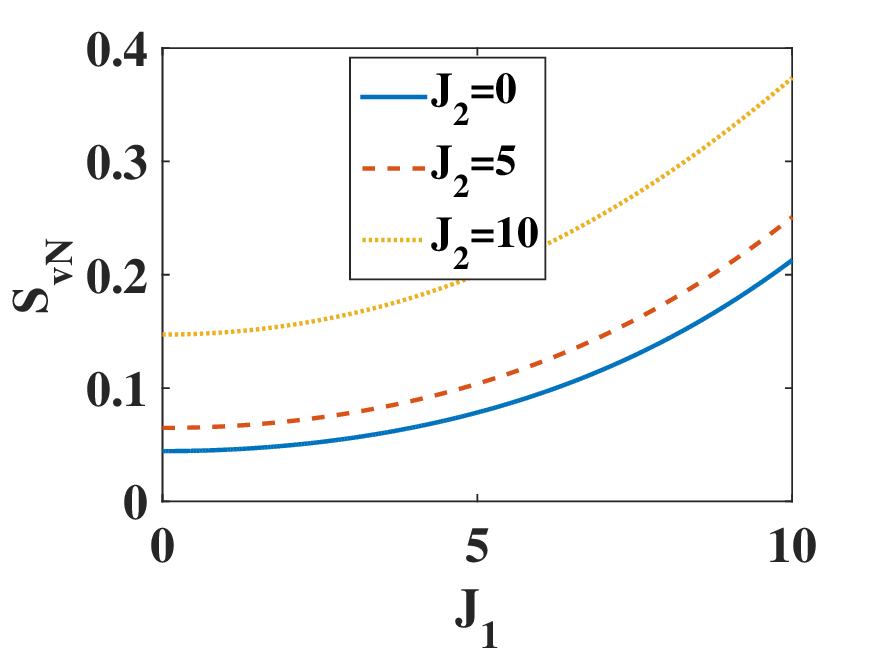}\\(a) & (b)
\end{array}$ 
\caption{(Color online)  Variation of (a) the efficiency $\eta$ and (b) the von Neumann entropy $S_{vN}$ as a function of $J_1$,  at the critical point $B_L=J_1/2$, for different values of $J_2$. The other parameters are the same as in Fig. \ref{Fig: QH_QL_W___vs___BL____P1}.}
\label{Fig: eta__S__vs___J1__diffJ2}
\end{figure}

\begin{figure}[!h]
 $\begin{array}{cc}
 \includegraphics[width=0.49\linewidth]{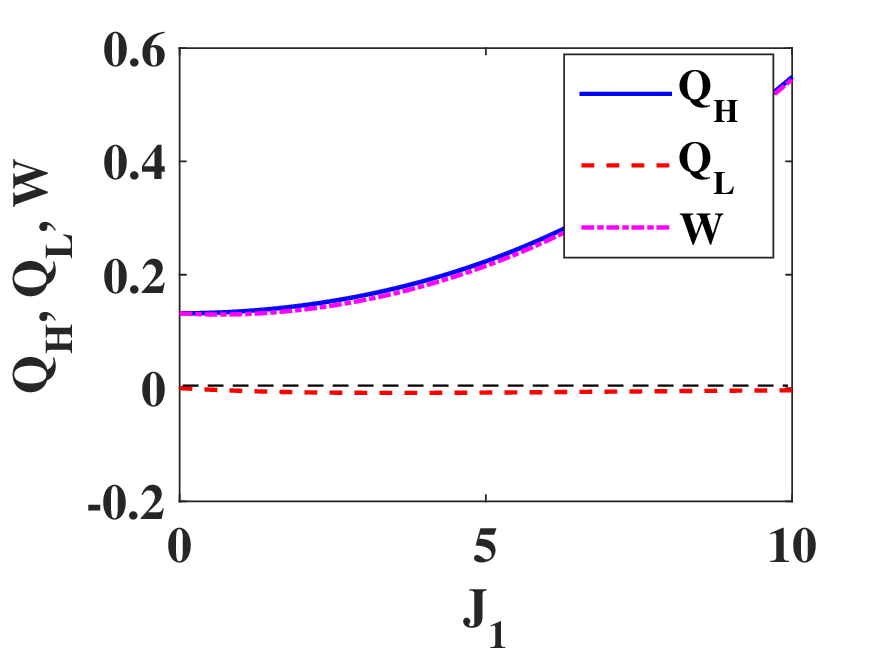}&
\includegraphics[width=0.49\linewidth]{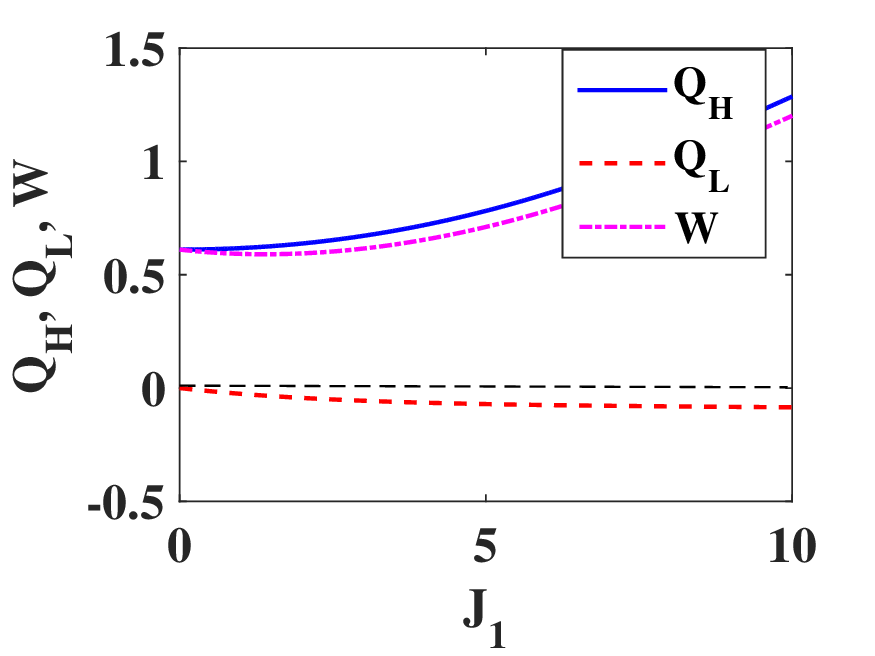}\\(a) & (b)
\end{array}$ 
\caption{(Color online) Variation of heat-exchanged $Q_H$ (solid blue) and $Q_L$ (dotted red), with the hot and the cold bath, respectively and the net work done $W$ (dot-dashed magenta) as a function of $J_1$, at the critical point $B_L=J_1/2$, for (a) $J_2=0$ and (b) $J_{2}=10$. The other parameters are the same as in Fig. \ref{Fig: QH_QL_W___vs___BL____P1}.}
\label{Fig: W__vs___J1__diffJ2}
\end{figure}

Next, we discuss how the system behaves {\it at the critical point}, $B_L=J_1/2$. In this domain, the efficiency varies with $J_1$ differently than in the domain above the critical point. For small $J_1$, the system behaves as weakly coupled two-ion system, and the efficiency of the engine becomes similar to that of a single-ion heat engine \cite{Our Paper- Single ion}. Note that for a single ion, driven by a magnetic field alone, the efficiency of an Otto engine can be written as $\eta=1-B_L/B_H$ \cite{Our Paper- Single ion}. Therefore, as $B_L$ is increased from zero, the efficiency decreases from close to unity, {\it linearly}. Clearly the effect of $J_1$ is not very substantial. For larger $J_1$, however, the above linear dependence does not hold any more. As $J_1$ becomes large, i.e., as two ions get coupled stronger, their internal correlation eventually increases the efficiency [see Fig. \ref{Fig: eta__S__vs___J1__diffJ2}(a)]. This also corresponds to an increase in the von Neumann entropy of the system [see Fig. \ref{Fig: eta__S__vs___J1__diffJ2}(b)]. More interestingly, the work $W$ done by the system varies with $J_1$ (Fig. \ref{Fig: W__vs___J1__diffJ2}) in the similar way the entropy varies.

Next we study the critical point behavior of the engine under the action of the ion 3.  It is known that at the critical point, the correlation becomes long range. Essentially we ask the following question: Does the correlation with the ion 3 have any substantial effect on the engine efficiency? The interaction of the system $S$ to the ion 3 is described by a direct coupling $H^{\rm int}_{23}$ and an indirect coupling $H_{3,{\rm ph}}$ via a common vibrational mode $a$. Such interaction affects the entropy of the system, in addition to the entropy change by a thermal environment.  

%

\begin{figure}[!h]
$\begin{array}{cc}
\includegraphics[width=0.49\linewidth]{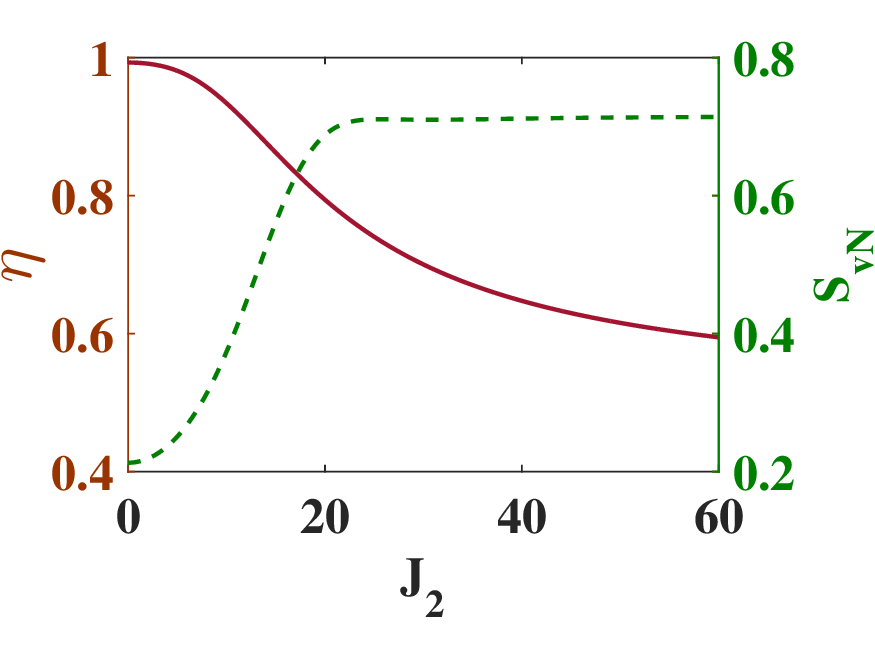}&
\includegraphics[width=0.49\linewidth]{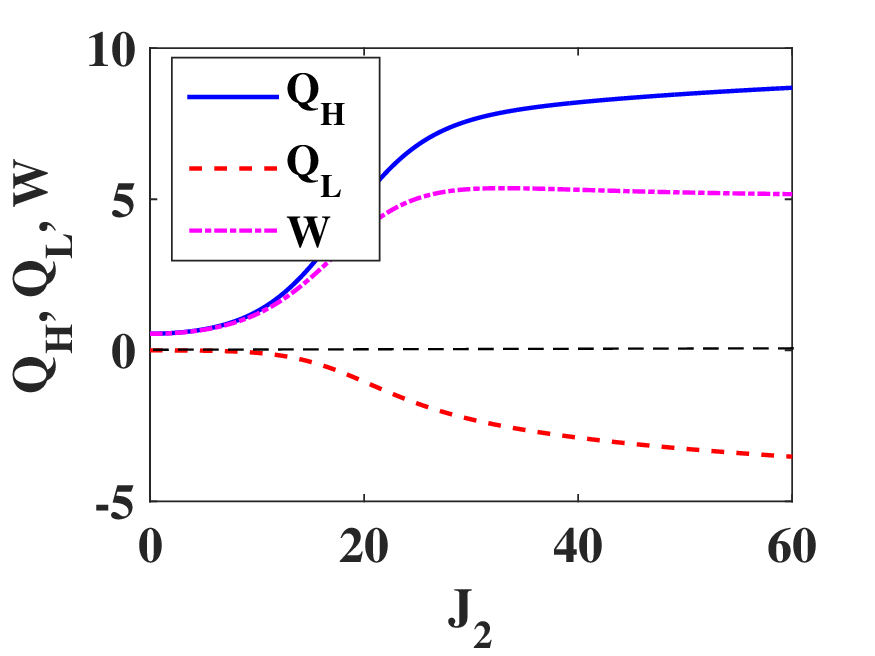}\\(a) & (b)
\end{array}$
\caption{(Color online)  Variation of (a) the efficiency $\eta$ (solid red) and von Neumann entropy $S_{vN}$ (dashed green) and (b) the heat exchanged $Q_H$ (solid blue) and $Q_L$ (dotted red), with the hot and the cold bath, respectively, and the net work done $W$ (dot-dashed magenta), as a function of $J_2$ at the critical point $B_L=J_1/2=5$. The other parameters are the same as in Fig. \ref{Fig: QH_QL_W___vs___BL____P1}. The system is measured in the $|E_1\rangle$  state.}
\label{Fig: eta__S__vs___J2___P1}
\end{figure}


When $J_2=0$ (no direct coupling to the ion 3), the $S_{vN}$ of the system $S$ is primarily affected by the heat exchange with the thermal bath. But when $J_2\neq 0$, the correlation with the ion 3 starts playing a major role in the engine operation. At the critical point ($B_L=J_1/2$), the von Neumann entropy $S_{vN}$ of the system $S$ increases with  $J_{2}$, before getting saturated [see Fig. \ref{Fig: eta__S__vs___J2___P1}(a)]. This is due to the direct interaction between  the system and the ancillary qubit (i.e., the ion 3), that leads to further mixing in the density matrix of $S$. This is akin to the increase in entropy when a system interacts with its thermal environment. However, such an increase in entropy with $J_2$ is not effective in increasing the efficiency $\eta$. We find that $\eta$ decreases monotonically with $J_2$ [see both Fig. \ref{Fig: eta__S__vs___J1__diffJ2} and Fig. \ref{Fig: eta__S__vs___J2___P1}(a)], though the work $W$ done by the system increases [see Fig. \ref{Fig: eta__S__vs___J2___P1}(b)]. This means that if the system $S$ is open to interaction with any ancillary system, it degrades the performance of the system as a heat engine, even if it delivers more work. 

We further note that the work $W$ and the entropy vary in the similar way with $J_2$. This marks an one-to-one correspondence between the work and the entropy. As mentioned before, such a correspondence is also found from the Fig. \ref{Fig: eta__S__vs___J1__diffJ2}(b) and Fig. \ref{Fig: W__vs___J1__diffJ2}. A linear dependence of the work output on the entropy has been reported also in heat engines based on molecular systems \cite{hubner}.


We emphasize that, as evident from the Figs. \ref{Fig: eta__S__vs___J1__diffJ2} and Fig. \ref{Fig: eta__S__vs___J2___P1}, there is no generic correlation between the von Neumann entropy and the efficiency of the engine. While the entropy of a system increases due to interaction among the constituent particles as well as due to external perturbation, its efficiency would primarily depend on the nature of this interaction. The internal interaction would improve the engine performance, while any interaction with an ancillary system deteriorates the system's performance as an efficient heat engine. This implies that when a part of an extended system is considered as the working fluid, the correlation of such working fluid with the other part tends to adversely affect its efficiency as a heat engine. However, a strong interaction between the particles of the working fluid would improve the efficiency of the system. 

On the other hand, if the state of the system becomes more mixed due to these interactions, then the system delivers more work. Both the internal and external interactions will allow the system to generate more work output, albeit with less efficiency if $J_2$ increases. This essentially means that the coherence in the system is consumed to deliver further work. If there is no further change in entropy (i.e., no further decrease in coherence content), work cannot be further generated. 

\begin{figure}[!h]
$\begin{array}{cc}
\includegraphics[width=0.49\linewidth]{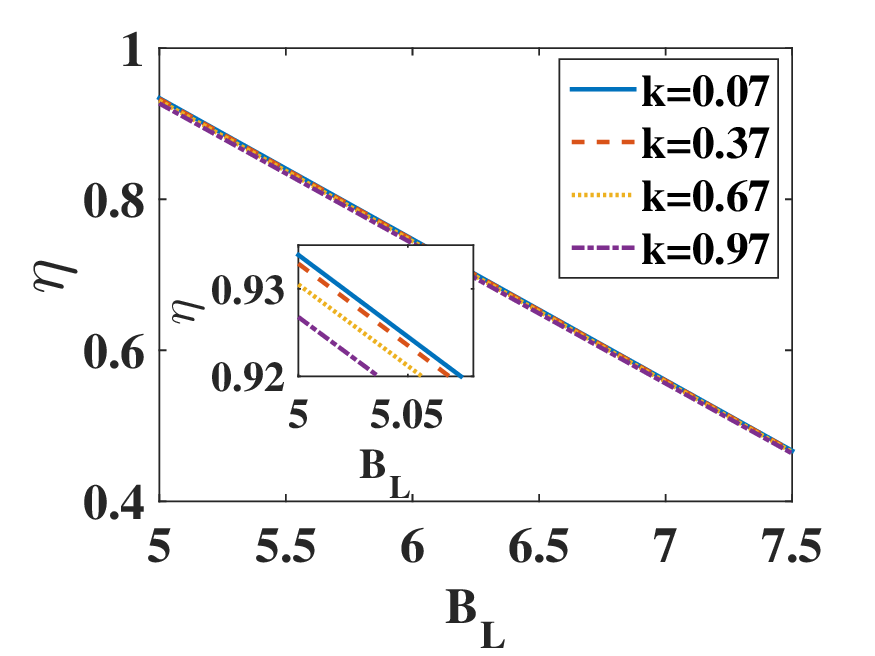}&
\includegraphics[width=0.49\linewidth]{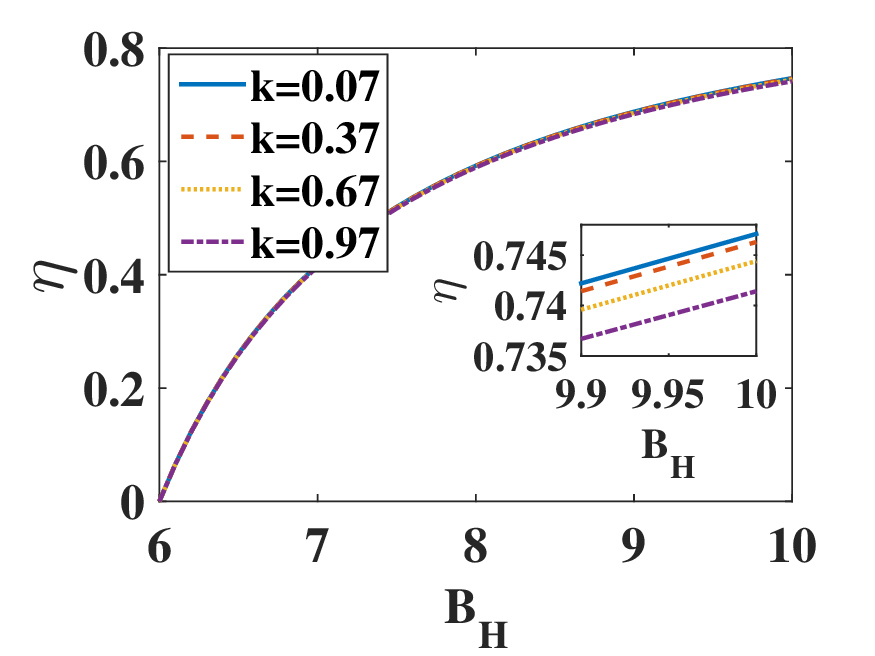}\\(a) & (b)
\end{array}$
\caption{Variation of the  efficiency $\eta$ as a function of (a) the magnetic field  $B_{L}$ and (b) the magnetic field  $B_{H}$ during isochoric heating step for different values of the coupling $k$ between the internal and motional states of the ions, while the system is measured in  $|E_1\rangle$ state. For (a), $B_{L}$ is maintained above the critical point, while for (b), we have chosen $B_{L}=6$, a value above the critical point. The other parameters are the same as in Fig. \ref{Fig: QH_QL_W___vs___BL____P1}. Clearly, changes in the coupling $k$ (within the Lamb-Dicke limit) have only negligible effect of the efficiency $\eta$ (as shown in the insets) }
\label{k-dependence}
\end{figure}

Note that the Hamiltonians $H_{12,{\rm ph}}$ and $H_{3,{\rm ph}}$ describe the interaction of the working medium (two ions) and the axillary system to the cold bath. For larger values of $k$ (stronger coupling to the bath), the system will move more out of equilibrium, and the performance of the engine would be degraded (see, e.g., Ref. \cite{strong-coupling}). However, in the Lamb-Dicke limit, for smaller values of $k$, the efficiency is not substantially affected, as displayed in Fig. \ref{k-dependence}.

 \subsection{Discussions}
 
In our model, after each cycle ends, the total change in entropy is zero, as the system is initialized to the ground state at the end of each cycle.  This is achieved as (a) the system is  projected into the ground state during the
cooling stage and (b) all the adiabatic stages are chosen to be perfect. Therefore, there exists no internal irreversibility in our model, unlike that in finite-time thermodynamics (FTT),
for which internal ``friction'' leads to finite change in entropy ($\Delta S \ne 0$) \cite{Endoreversible Thermodynamics}. This means that there are no heat leaks involved, as there are no changes in the probabilities of the eigenstates during these stages. In fact, during the adiabatic stages (when $\tau\ll1/\gamma$, i.e., as long as the interaction with the external bath does not become effective), the system
evolves reversibly (i.e., unitarily, in quantum mechanical terminology, see  Ref. \cite{Quan-Thermodynamics cycle} for the equivalence).  Interestingly, these stages occur for finite time $\tau$, which is long enough to ensure adiabaticity, and yet fast enough to ensure that the bath interaction is not set in.  Only when the system is coupled to the bath, it undergoes an irreversible process (thermalization or projective measurement).  Therefore, our model can be considered as endoreversible \cite{Endoreversible Thermodynamics}, though not FTT (the thermalization process in our model is infinitely slow).  Note that heat leaks happen in Curzon-Ahlborn model for its fast operation, while the internal irreversibilities occur due to entropy change during the cycle (see Sec. 5, Ref. \cite{Endoreversible Thermodynamics}). We emphasize that neither of them occurs in our case.

  
  This proposal can be implemented using the current trapped-ion technology. We have considered the parameter regime as described in  Ref. \cite{Measuring the heat exchange of quantum process}. They proposed a scheme based on two laser-cooled trapped $^{40}$Ca$^+$ ions of mass $m$ confined in a harmonic potential. An external magnetic field gives rise to a Zeeman splitting, typically in the
range of $2\pi\times5$ MHz to $2\pi\times20$ MHz.  The trap frequency $\omega$ is typically in the range of $2\pi\times1$ MHz to $2\pi\times5$ MHz. The temperature can be varied from 2 mK to below 6 $\mu$K. The Lamb-Dicke parameter, proportional to $k$ governs the coupling strength between internal and motional states. Typical value of $k$ can be 0.07, whereas the coupling $J_i$s between the internal states of the ions for a typical ion trap system can be $2\pi\times1.5$ kHz \cite{Quantum dynamics of trapped ions in a dynamic field gradient using dressed states}.

\section{Conclusion}\label{s:v}
In summary, we have studied how an Otto engine with two trapped ions behaves at and in the neighbourhood of the critical point of quantum phase transitions. While the system behaves as a heat engine above the critical point, it fails to act as one below this point. This suggests that the engine efficiency can be considered as a marker for identifying the phase transition. Similar conclusions can be made when the magnetic field is reversed and the system behaves as a heat engine above the critical point. Though the adiabatic decrease in the magnetic field increases the efficiency of the heat engine, the lower limit of the magnetic field (corresponding to the critical point) is governed by the internal coupling constant of the constituent ions of the system $S$.   

Further, at the critical point, the coupling of the system $S$ with another ion modifies the performance of the QOE. It is found that as the internal coupling inside the system $S$ is increased, the efficiency gets enhanced. On the other hand, on increase of the coupling to an ancillary system (here, the ion 3), the efficiency decreases. We therefore conjecture that if the working fluid for a quantum heat engine is a part of a larger system, then its interaction with the other part of the system (or the long-range correlation with the ancillary system) degrades the overall performance of the working fluid as a heat engine, though more work could be extracted from the system in such cases. 

\section*{Acknowledgments}
 We extend our sincere thanks to the reviewers, whose insightful comments have led to substantial improvement of the paper.
 
\appendix*
\section{}
The linear behavior in Fig. 3(b) can be understood as follows. This plot is obtained by changing $B_L$ (for a fixed $B_H$ and $J_1$), and calculating $W$ and $\eta$ for each value of $B_L$. In the limit of very weak coupling to the photon modes (i.e., for negligible $k$) and $J_2=0$, we find that the work output can be written as $W=(B_H-B_L)f_1(J_1,B_H)$, where 
\begin{eqnarray}
f_{1}(J_{1},B_{H}) &=& 1-\frac{1}{Z}\sinh\left(\frac{2B_H}{k_BT_H}\right)\,,\nonumber \\ Z &=&\cosh\left(\frac{2B_H}{k_BT_H}\right)+\cosh\left(\frac{J_1}{k_BT_H}\right)\;.\,\,\,\,
\label{app1}
\end{eqnarray}
Clearly for $B_L<B_H$, the work output is positive for all positive values of $J_1$ and $B_H$. We further find that the efficiency becomes  
\begin{eqnarray}
\eta &=&\left(1-\frac{B_L}{B_H}\right)\frac{f_1}{f_1-f_2}\,,\nonumber \\f_{2}\left(J_{1},B_{H}\right)&=&\frac{J_{1}}{2B_{H}}\frac{1}{Z}\sinh\left(\frac{J_1}{k_BT_H}\right)\;.
\label{app2}
\end{eqnarray}

Clearly, in the limit of $J_1\rightarrow 0$ (i.e., when two ions do not interact with each other), $\eta=1-B_L/B_H$, that corresponds to a single-ion heat engine. Further note that the magnetic field cannot be reduced below $J_1/2$ (the critical point) to make the heat engine work, i.e., $B_L$ must be at least $J_1/2$.  This means that $B_H$ needs to be chosen greater than $J_1/2$ for the system to behave as a heat engine, as $B_H>B_L$ is required to have a positive work output.

More importantly, as seen from the Eqs. (\ref{app1}) and (\ref{app2}),  both $W$ and $\eta$ decrease linearly with increase in $B_L$ and therefore $\eta$ becomes proportional to $W$ ($\eta=W/\{B_H(f_1-f_2)\}$) for fixed values of $B_H$ and  $J_1$. This explains the linear behavior of Fig. 3(b). There is no extremum in this case, as both $dW/dB_L$ and $d\eta/dB_L$ are finite, if $B_H$ is finite. Therefore, the efficiency does not follow $(1/2)(1-B_L/B_H)$ behavior, unlike that as described in Ref. \cite{uzdin}.  It must be borne in mind that the achievable upper limit of $\eta$ corresponds to the minimum allowed value of $B_L=J_1/2$ (the critical point), below which the system does not work as a heat engine. This means that the efficiency cannot be {\it arbitrarily} increased to unity. 

\begin{figure}[!h]
$\begin{array}{cc}
\includegraphics[width=0.49\linewidth]{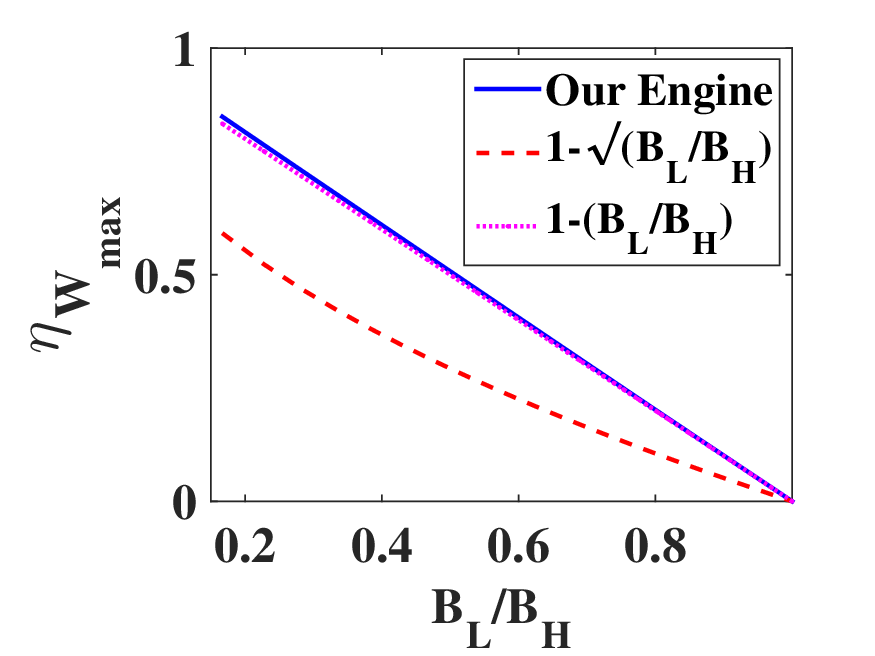}&
\includegraphics[width=0.49\linewidth]{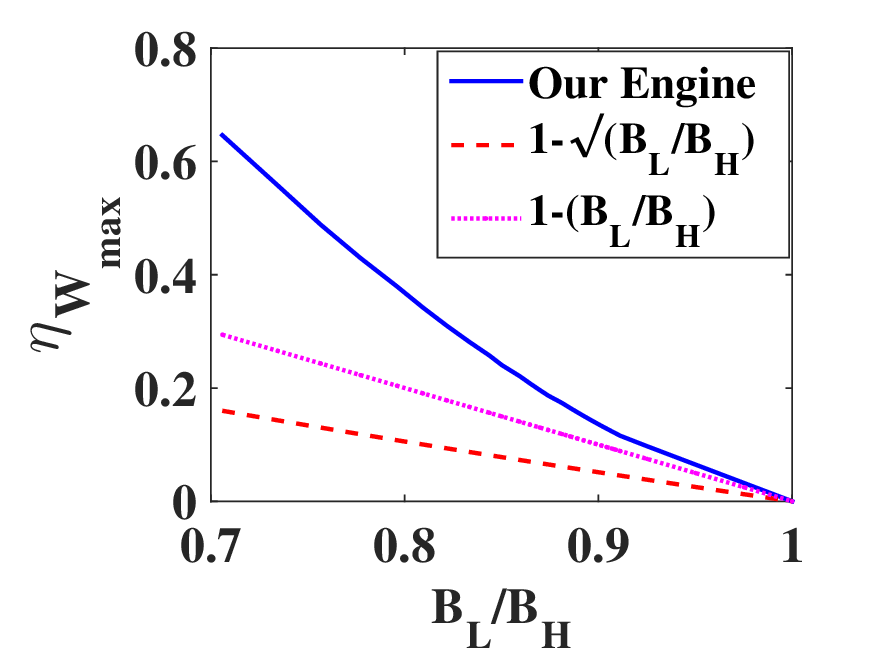}\\(a) & (b)
\end{array}$
\caption{Variation of $\eta_{W_{max}}$ in our model (solid blue) with $\frac{B_{L}}{B_{H}}$ for (a) $J_{1}=1$ and (b) $J_{1}=10$, while the system is measured in $\left|E_{1}\right\rangle $ state.  The plots for $\left(1-\sqrt{\frac{B_{L}}{B_{H}}}\right)$ (dashed red) and $\left(1-\frac{B_{L}}{B_{H}}\right)$ (dotted magenta)  are also shown for reference. We have chosen values of $B_{L}$ above the critical point. We have varied $B_{L}$ from 0.7 to $B_{H}=10$  in (a) and from 6 to $B_{H}=20$ in (b).  The other parameters are the same as in Fig. \ref{Fig: QH_QL_W___vs___BL____P1}.}
\label{eta_W_max_above_critical_point}
\end{figure}

The Fig. 4(b) is obtained by calculating $W$ and $\eta$ at different values of $B_H$ (for a fixed $J_1$ and $B_L>J_1/2$, i.e., above critical point). As seen from Eqs. (\ref{app1}) and (\ref{app2}), both $W$ and $\eta$ vary in a nontrivial way with $B_H$.  One would obtain a maximum of $W$ for a certain value of $B_H$, which could be obtained by using $\partial W/\partial B_H=0$. We calculated  $\eta$ using this particular value of $B_H$, for which  $W$ reaches to a maximum value  $W_{max}$. The above procedure is repeated for other values of $B_L$, so as to investigate how $\eta_{W_{max}}$ varies with $B_L/B_H$. We show in Fig. \ref{eta_W_max_above_critical_point} the variation of $\eta_{W_{max}}$ with $\frac{B_{L}}{B_{H}}$ for different values of $J_1$. This clearly shows that for smaller $J_{1}$, the efficiency $\eta_{W_{max}}$ corresponding to maximum achievable work, displays a behavior close to $1-\frac{B_{L}}{B_{H}}$ as shown in Fig. \ref{eta_W_max_above_critical_point}(a). On the other hand, for larger $J_1$, this behavior deviates far from $1-\frac{B_{L}}{B_{H}}$. This further suggests that the ion-ion interaction enhances both the efficiencies $\eta$ and $\eta_{W_{max}}$. 

We emphasize that though the engine described in this work exhibits endoreversible cycle, its efficiency does not follow a  ($1-\sqrt{B_L/B_H}$) behavior, that would resemble that for an classical endoreversible heat engine (with magnetic field terms replaced by ambient temperature terms) \cite{Endoreversible Thermodynamics,ouerdane}. The crucial differences lie into Markovian heating process (unlike polynomial heat laws, often used in case of classical endoreversible engines; e.g., see Sec. 5.2, \cite{Endoreversible Thermodynamics}) and the measurement-based cooling protocol, which does not have any classical analog.


\begin{thebibliography}{50}

\bibitem{Kerson Huang book on Statistical Mechanics} K. Huang, {\it Statistical Mechanics}, 2nd ed. (John Wiley \& Sons, New York, 1987).

\bibitem {Scovil maser paper} H. E. D. Scovil and E. O. Schulz-DuBois, Phys. Rev. Lett. {\bf 2}, 262 (1959).


\bibitem{Thermodynamics book G. Mahler} G. Gemma, M. Michel, and G. Mahler, { \it Quantum Thermodynamics} (Springer, New York, 2004).

\bibitem{The Role of Quantum Information in Thermodynamics—A Topical Review-J. Goold}  J. Goold, M. Huber, A. Riera, L. del Rio, and P. Skrzypczyk, J. Phys. A {\bf 49}, 143001 (2016).

\bibitem{Quantum Thermodynamics-S. Vinjanampathy}  S. Vinjanampathy and J. Anders, Contemp. Phys. {\bf 57}, 545 (2016).

\bibitem{Introduction to Quantum Thermodynamics: History and Prospect- R. Kosloff}  R. Alicki and R. Kosloff, arXiv:1801.08314 (2018).


\bibitem{Quan-Thermodynamics cycle} H. T. Quan, Y.-x. Liu, C. P. Sun, and F. Nori, Phys. Rev. E {\bf 76}, 031105 (2007); H. T. Quan, Phys. Rev. E {\bf 79}, 041129 (2009).


\bibitem{kosloff} T. Feldmann and R. Kosloff, Phys. Rev. E  {\bf 61}, 4774 (2000); {\bf 70}, 046110 (2004).

\bibitem{Effects of reservoir squeezing on quantum systems and work extraction} X. L. Huang, T. Wang, and X. X. Yi, Phys. Rev. E {\bf 86}, 051105 (2012).

\bibitem{Quantum Otto engine of a two-level atom with single-mode fields} J. Wang, Z. Wu, and J. He, Phys. Rev. E {\bf 85}, 041148 (2012).

\bibitem{Four-level entangled quantum heat engines} T. Zhang, W.-T. Liu, P.-X. Chen, and C.-Z. Li, Phys. Rev. A {\bf 75}, 062102 (2007).


\bibitem{Entangled QHE based on two two spin systems with DM} G.-F. Zhang, Eur. Phys. J. D {\bf 49}, 123 (2008).

\bibitem{Thermal entanglement in two-atom cavity QED and the entangled quantum Otto engine} H. Wang, S. Liu, and J. He, Phys. Rev. E {\bf 79}, 041113 (2009).

\bibitem{johal}G. Thomas and R. S. Johal, Phys. Rev. E {\bf 83}, 031135 (2011).

\bibitem{Atlintas} F. Atlintas, A. \"U. C. Hardal, and \"O.  E. M\"ustecapl\ifmmode \imath \else \i \fi{}o\ifmmode \breve{g}\else \u{g}\fi{}lu, Phys. Rev. E {\bf 90}, 032102 (2014); F. Atlintas and  \"O.  E. M\"ustecapl\ifmmode \imath \else \i \fi{}o\ifmmode \breve{g}\else \u{g}\fi{}lu, Phys. Rev. E {\bf 92}, 022142 (2015).

\bibitem{Thermal entangled four level QOE} H. Xian and H. JiZuou, Sci. China-Phys. Mech. Astron., {\bf 55}, 1751 (2012).

\bibitem{A Special entangled QHE based on two qubit Heisenberg XX model} X. L. Huang, Huan Xu, X. Y. Niu, and Y. D. Fu, Phys. Scr. {\bf 88}, 065008 (2013).

\bibitem{Special coupled quantum otto cycle} H. L. Hung, Y. Liu, Z. Wang, and X. Y. Niu, Eur. Phys. J. Plus {\bf 129}, 4 (2014).

\bibitem{Thermal entangled QHE working with 3-qubit XX model} J.-Z. He, X. He, and J. Zheng, Int. J. Theor. Phys. {\bf 51}, 2066 (2012).

\bibitem{Lipkin-Meshkov-Glick Model in QOE} S. Cakmak, F. Altintas, and O. Mustecaphioglu, Eur. Phys. J. Plus {\bf 131}, 197 (2016). 

\bibitem{Performence of coupled system as QHE} G. Thomas, M. Banik, and S. Ghosh, Entropy {\bf 19}, 442 (2017).




\bibitem{Quantum harmonic oscillator Kosloff} Y. Rezek and R. Kosloff, New J. Phys. {\bf 8}, 83 (2006); R. Kosloff and Y. Rezek,  Entropy {\bf 19}, 136 (2017).

\bibitem{Efficiency at maximum power of a quantum heat engine based on two coupled oscillators} J. Wang, Z. Ye, Y. Lai, W. Li, and J. He, Phys. Rev. E {\bf 91}, 062134 (2015).


\bibitem{Quantum-classical transition of photon-Carnot engine induced by quantum decoherence-Quan} H. T. Quan, P. Zhang, and C.P Sun, Phys. Rev. E {\bf 73}, 036122 (2006).

\bibitem{single ion-Abah} O. Abah, J. Ro\ss nagel, G. Jacob, S. Diffner, F. Schmidekaler, K. Singer, and E. Lutz, Phys. Rev. Lett. {\bf 109}, 203006 (2012);  J. Ro\ss Ÿnagel, O. Abah, F. Schmidt-Kaler, K. Singer, and E. Lutz, Phys. Rev. Lett. {\bf 112}, 030602 (2014).

\bibitem{optomechanical system-Meystre} K. Zhang, F. Bariani, and P. Meystre, Phys. Rev. Lett. {\bf 112}, 150602 (2014); Phys. Rev. A {\bf 90}, 023819 (2014).

\bibitem{Magnon-driven quantum-dot heat engine} B. Sothmann and M. Buttiker, Europhys. Lett. {\bf 99}, 27001 (2012).

\bibitem{Isolated quantum heat engine- bosonic system} O. Fialko and D. W. Hallwood, Phys. Rev. Lett. {\bf 108}, 085303 (2012).

\bibitem{Scully Science}M. O. Scully, M. S. Zubairy, G. S. Agarwal, and H. Walther, Science {\bf 299}, 862 (2003); M. O. Scully, Phys. Rev. Lett. {\bf 88}, 050602 (2002); M. O. Scully, Phys. Rev. Lett. {\bf 87}, 220601 (2001).

\bibitem {Quantum Brayton cycle with coupled systems as working substance} X. L. Huang, L.C. Wang, and X.X. Yi, Phys. Rev. E {\bf 87}, 012144 (2013).















\bibitem{S. Sachdev book- Quantum Phase Transition} S. Sachdev, {\it Quantum Phase Transition} (Cambridge University Press, Cambridge, 1999).

\bibitem{Quantum thermodynamic cycle with quantum phase transition} Y.-Han Ma, S.-He Su, and C.-Pu Sun, Phys. Rev. E {\bf 96}, 022143 (2017).

\bibitem{blatt}D. Leibfried, R. Blatt, C. Monroe, and D. Wineland, Rev. Mod. Phys. {\bf 75}, 281 (2003).

\bibitem{Monroe ion trap}  C. Monroe, D. M. Meekhof, B. E. King, S. R. Jefferts, W. M.
Itano, D. J. Wineland, and P. Gould, Phys. Rev Lett. {\bf 75}, 4011 (1995).

\bibitem{Mam and Brummer paper} A. Biswas and P. Brumer, Israel J. Chem.  {\bf 52}, 461 (2012).

\bibitem{strong-coupling} D. Newman, F. Mintert, and A. Nazir, Phys. Rev. E {\bf 95}, 032139 (2017).



\bibitem{Single-Atom Heat Machines Enabled by Energy Quantization} D. Gelbwaser-Klimovsky, A. Bylinskii, D. Gangloff, R. Islam, A. Aspuru-Guzik, and V. Vuletic,, Phys. Rev. Lett. {\bf 120}, 170601 (2018).

\bibitem{Quantum four-stroke heat engine Thermodynamics observable in a model with intrinsic friction}T. Feldmann and R. Kosloff, Phys. Rev. E {\bf 68}, 016101 (2003).

\bibitem{Irreversible Work and Inner Friction in Quantum Thermodynamics Process} F. Plastina {\it et al.}, Phys. Rev. Lett. {\bf 113}, 260601 (2014).

\bibitem{Friction due to inhomogeneous driving of coupled spins in a quantum heat engine}G. Thomas and R. S. Johal, Eur. Phys. J. B {\bf 87}, 166 (2014).

\bibitem{Irreversible work and internal friction in a quantum Otto cycle of a single arbitrary spin} S. Cakmak, F. Altintas, A. Gencten, and \"O. E. M\"ustecapl\ifmmode \imath \else \i \fi{}o\ifmmode \breve{g}\else \u{g}\fi{}lu, Eur. Phys. J. D {\bf 71}, 75 (2017).

\bibitem{Our Paper- Single ion} S. Chand and A. Biswas, Europhys. Lett. {\bf 118}, 6 (2017).

\bibitem{Our Paper- Two ions} S. Chand and A. Biswas, Phys. Rev. E {\bf 95}, 032111 (2017).












\bibitem{Work and quantum phase transitions: Quantum latency} E. Mascarenhas, H. Braganca, R. Dorner, M. Franca Santos, V. Vedral, K. Modi, and J. Goold, Phys. Rev. E {\bf 89}, 062103 (2014).


\bibitem{Quantum resources for purification and cooling: fundamental limits and opportunities}F. Ticozzi and L. Viola, Sci. Rep. {\bf 4}, 5192 (2014).

\bibitem{Cooling in strongly correlated optical lattices: prospects and challenges} D. McKay and B. DeMarco, Rep. Prog. Phys. {\bf 74}, 054401 (2011).

\bibitem{The power of a critical heat engine} M. Campisi and R. Fazio, Nature Commun. {\bf 7}, 11895 (2016).

\bibitem{Generalized model and optimum performance of an irreversible quantum brayton engine with spin system}F. Wu, L. Chen, F. Sun, and Q. Li, Phys. Rev. E {\bf 73}, 016103 (2006).

\bibitem{gordon}J. M. Gordon, Am. J. Phys. {\bf 59}, 551 (1991).

\bibitem{Von Neumann book}John von Neumann. {\it Mathematical Foundations of Quantum Mechanics}, (Princeton University Press, Princeton, NJ, 1955).

\bibitem{hubner} W. Hubner, G. Lefkidis, C. D. Dong, D. Chaudhuri, L. Chotorlishvili, and J. Berakdar, Phys. Rev. B {\bf 90}, 024401 (2014).

\bibitem{Endoreversible Thermodynamics}K. H. Hoffmann, J. Burzler, and S. Schubert, J. Nonequilib. Thermodyn. {\bf 22}, 311 (1997).

\bibitem{Measuring the heat exchange of quantum process} J. Goold {\it et al.}, Phys. Rev. E {\bf 90}, 020101(R) (2014).

\bibitem{Quantum dynamics of trapped ions in a dynamic field gradient using dressed states} S. W\"olk and C. Wunderlich, New. J. Phys. {\bf 19}, 083021 (2017).

\bibitem{uzdin}R. Uzdin and R. Kosloff, Europhys. Lett. {\bf 108}, 40001 (2014).

\bibitem{ouerdane} H. Ouerdane, Y. Apertet, C. Goupil, and Ph. Lecoeur,, Eur. Phys. J. Special Topics {\bf 224}, 839 (2015). 


\end{thebibliography}
\end{document}